\documentclass[journal=nalefd,manuscript=article]{achemso}

	\usepackage{achemso}
	\usepackage{bm}
	\usepackage{braket}
	\usepackage{dsfont}
\usepackage[dvipsnames]{xcolor}
\definecolor{mypink2}{RGB}{255,0,255}
 	\usepackage{graphicx}
	\usepackage[caption=false]{subfig}
	\usepackage[export]{adjustbox}
	\usepackage{amsmath}
	\usepackage{amssymb}
	\usepackage{xcolor}
\usepackage{setspace}

	\usepackage{hyperref}		
	\usepackage[figure]{hypcap}		

	\hypersetup{
	   	bookmarks		= false,		
		unicode			= false,	
		pdfborder		= {0 0 0}	
		pdftoolbar		= false,		
		pdfmenubar		= true,		
		pdffitwindow		= false,     	
		pdfstartview		= {FitH},    	
		pdfnewwindow	= true,      	
		colorlinks		= true,		
		linkcolor			= black,		
		citecolor			= black,		
		filecolor			= black,		
		urlcolor			= black,		
		linktocpage		= true,		
		pdftitle			= {Moir\'{e} superlattice effects and band structure evolution in near-30-degree twisted bilayer graphene},
		pdfauthor		= {M. Mucha-Kruczynski},
		pdfsubject		= {},
		pdfcreator		= {MikTeX},
		pdfproducer		= {Marcin Mucha-Kruczynski},
		pdfkeywords		= {},
		}

	\newcommand{\vect}[1]{\boldsymbol{#1}}		
	
	\newcommand{\comment}[1]{\ignorespaces}         

\title{Moir\'{e} Superlattice Effects and Band Structure Evolution in Near-30-Degree Twisted Bilayer Graphene}
\author{Matthew J.~Hamer}
\affiliation{Department of Physics, University of Manchester, Oxford Road, Manchester, M13 9PL, United Kingdom}
\alsoaffiliation{National Graphene Institute, University of Manchester, Booth Street East, Manchester, M13 9PL, United Kingdom}
\author{Alessio Giampietri}
\affiliation{Elettra-Sincrotrone Trieste ScPA, Trieste 34149, Italy}
\author{Viktor Kandyba}
\affiliation{Elettra-Sincrotrone Trieste ScPA, Trieste 34149, Italy}
\author{Francesca Genuzio}
\affiliation{Elettra-Sincrotrone Trieste ScPA, Trieste 34149, Italy}
\author{Tevfik O.~Mente\c{s}}
\affiliation{Elettra-Sincrotrone Trieste ScPA, Trieste 34149, Italy}
\author{Andrea Locatelli}
\affiliation{Elettra-Sincrotrone Trieste ScPA, Trieste 34149, Italy}
\author{Roman V.~Gorbachev}
\affiliation{Department of Physics, University of Manchester, Oxford Road, Manchester, M13 9PL, United Kingdom}
\alsoaffiliation{National Graphene Institute, University of Manchester, Booth Street East, Manchester, M13 9PL, United Kingdom}
\alsoaffiliation{Henry Royce Institute, Oxford Road, Manchester, M13 9PL, United Kingdom}
\author{Alexei Barinov}
\email{Alexey.Barinov@elettra.eu}
\affiliation{Elettra-Sincrotrone Trieste ScPA, Trieste 34149, Italy}
\author{Marcin Mucha-Kruczy\'{n}ski}
\email{M.Mucha-Kruczynski@bath.ac.uk}
\affiliation{Department of Physics, University of Bath, Claverton Down, Bath BA2 7AY, United Kingdom}
\alsoaffiliation{Centre for Nanoscience and Nanotechnology, University of Bath, Claverton Down, Bath BA2 7AY, United Kingdom}


\begin{document}

\begin{abstract}
In stacks of two-dimensional crystals, mismatch of their lattice constants and misalignment of crystallographic axes lead to formation of moir\'{e} patterns. We show that moir\'{e} superlattice effects persist in twisted bilayer graphene (tBLG) with large twists and short moir\'{e} periods. Using angle-resolved photoemission, we observe dramatic changes in valence band topology across large regions of the Brillouin zone, including the vicinity of the saddle point at $\vect{M}$ and across 3 eV from the Dirac points. In this energy range, we resolve several moir\'{e} minibands and detect signatures of secondary Dirac points in the reconstructed dispersions. For twists $\theta>21.8^{\circ}$, the low-energy minigaps are not due to cone anti-crossing as is the case at smaller twist angles but rather due to moir\'{e} scattering of electrons in one graphene layer on the potential of the other which generates intervalley coupling. Our work demonstrates robustness of mechanisms which enable engineering of electronic dispersions of stacks of two-dimensional crystals by tuning the interface twist angles. It also shows that large-angle tBLG hosts electronic minigaps and van Hove singularities of different origin which, given recent progress in extreme doping of graphene, could be explored experimentally.

\end{abstract}

\maketitle

\newpage
  
Twisted bilayer graphene (tBLG) is the archetype of van der Waals heterostructures -- stacks of atomically thin materials with no directional bonding between consecutive layers and hence complete freedom of their relative rotational arrangement \cite{geim_nature_2013, novoselov_science_2016}. Tuning the twist angle, $\theta$, between lattice directions of neighboring crystals leads to formation of moir\'{e} superlattices (mSL), represented visually by patterns observed, for example, with scanning probe techniques \cite{xue_naturemat_2011, kang_nanoletters_2013, zhang_sciadv_2017}, and spatial modulation of interlayer coupling. This enables engineering of properties of a stack by tuning its stacking geometry, with examples including the observation of Hofstadter's butterfly \cite{ponomarenko_nature_2013, dean_nature_2013} and interfacial polarons \cite{chen_nanoletters_2018} in graphene/hexagonal boron nitride heterostructures, as well as interlayer excitons in transition metal dichalcogenide bilayers \cite{fang_pnas_2014, rivera_naturenano_2018}. In tBLG, at small angles, $\theta\approx 1^{\circ}$, mSL generate flat bands which host correlated electronic behavior including superconductivity \cite{cao_nature_2018, cao_nature_2018_2}. At the maximum twist angle, $\theta=30^{\circ}$, because the height-to-width ratio of a regular hexagon involves the irrational $\sqrt{3}$, tBLG is a quasicrystal \cite{ahn_science_2018, yao_pnas_2018}. However, properties of tBLG with twist angles between these two limits remain relatively unexplored experimentally, with the current studies mainly focused on the van Hove singularity due to hybridization of Dirac cone crossings \cite{li_naturephys_2010, ohta_prl_2012, brihuega_prl_2012, razado-colambo_scirep_2016, liao_nanolett_2015} which can be tuned with electric fields \cite{yu_prb_2019, jones_advmat_2020} and influences the optical properties of the stack \cite{patel_nanoletters_2015, yin_naturecomms_2016, yu_prb_2019}.
 
Here, we use angle-resolved photoemission spectroscopy (ARPES) to study evolution of the valence band structure of tBLG with large twist angles, $\theta\gtrsim 22^{\circ}$. We observe extensive modifications of the band structure not only near the intersections of the bands of the individual layers, but across a wide range of energies, $\sim 3$ eV, away from the Dirac points: appearance of multiple minigaps, signatures of additional Dirac points appearing in the dispersion and hybridisation of the isotropic bottoms of the graphene $\pi$-bands. We explain how these changes arise due to the coupling between the layers and mSL effects which persist at large twists when the apparent moir\'{e} wavelength is comparable to, but yet incommensurate with, the graphene lattice constant, and hence result in intervalley coupling. Our results demonstrate how, in a stack of two-dimensional crystals, the twist angle at an interface between two layers can be used to modify the electronic dispersion of the structure through a variety of mechanisms across a large range of $\theta$. Moreover, given the successful extreme doping of monolayer graphene close to and past its $\vect{M}$ van Hove singularity \cite{mcchesney_prl_2010, link_prb_2019, rosenzweig_prb_2019, rosenzweig_prl_2020}, the richness of the band structure we observe suggests large-twist tBLG as a playground to explore interplay of interaction effects driven by van Hove singularities.

\section{Results and Discussion}

We fabricated three tBLG devices, A, B and C, on top of hexagonal boron nitride ($h$-BN) using exfoliation and dry peel stamp transfer technique \cite{frisenda_csr_2018}. The tBLG samples were characterized by low-energy electron microscopy (LEEM) and low-energy electron diffraction (LEED) in order to determine the twist angles, $\theta= 22.6^{\circ}$ (tBLG-A), $26.5^{\circ}$ (tBLG-B) and $29.7^{\circ}$ (tBLG-C). For large twist angles, using reciprocal space LEED patterns to measure the twist is more precise than investigating the real space moir\'{e} periodicity with the scanning probe techniques (widely used for small $\theta$) because unit vectors for the latter are small, {\it i.e.} comparable with the graphene lattice constant. Our procedure, described in detail in the Supporting Information (SI), allows us to determine $\theta$ with the accuracy of $0.1^{\circ}$. In turn, by comparing the widths of the zeroth and first order LEED spots, we estimate the maximum twist angle disorder as $\Delta\theta=0.2^{\circ}$. This indicates relative homogeneity of the twist angle across areas of our devices much larger than the nano-ARPES spot size ($\lesssim 1$ $\mathrm{\mu}$m in diameter). The bottom graphene layer is rotated by $\theta_{h-\mathrm{BN}}\approx 10^{\circ}$, $15^{\circ}$ and $4^{\circ}$ for the A, B and C devices, respectively, with respect to the underlying $h$-BN -- this is sufficient to avoid moir\'{e} effects at the $h$-BN/graphene interface which are the strongest in highly-aligned $h$-BN/graphene structures \cite{ponomarenko_nature_2013, dean_nature_2013} and decrease with increasing $\theta_{h-\mathrm{BN}}$ \cite{jung_prb_2017}. All the measurements were performed at the Elettra Synchrotron and details of the fabrication process and discussion of the LEEM, LEED and ARPES experiments are provided in the SI.

\begin{figure*}[t]
	     \centering
	     \includegraphics[width=0.95\linewidth]{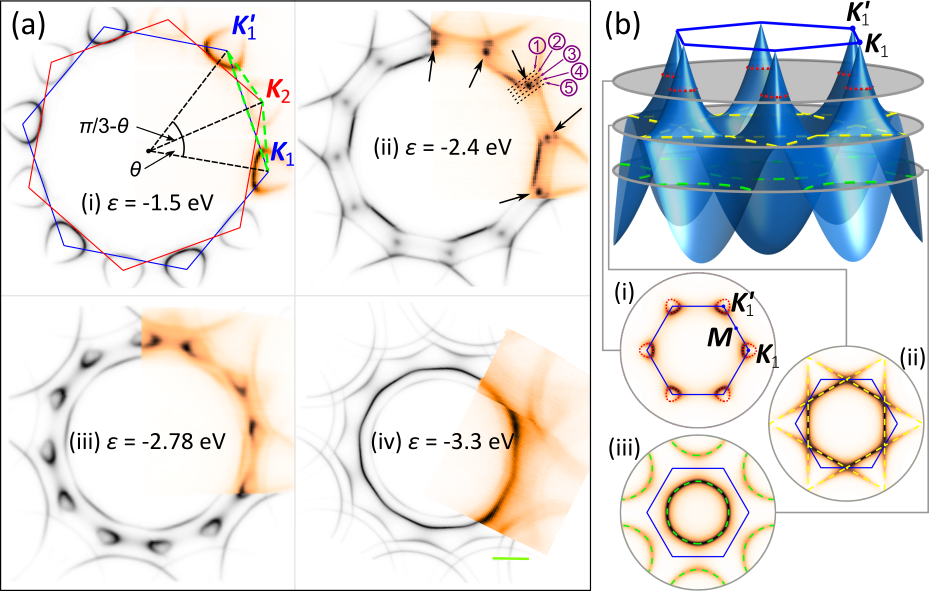}
	     \caption{Topology of tBLG energy contours. (a) ARPES constant-energy maps of tBLG-C, $\theta=29.7^{\circ}$; experimental data is shown in colour and theoretical simulation in black and white. The blue and red hexagons show Brillouin zones of the top ($i=1$) and bottom ($i=2$) layers, respectively, and the green dashed line indicates the $k$-space path for cuts in Fig.~\ref{fig:dos_and_gaps}(b). $\vect{K}_{i}$ and $\vect{K}^{'}_{i}$ denote inequivalent Brillouin zone corners in layer $i$. Black arrows in panel (ii) point to secondary Dirac points; dashed black line segments numbered with purple numbers show paths of cuts presented in Fig.~\ref{fig:sDP}. All panels show the same $k$-space area; the green scale bar in (iv) corresponds to 0.5 \AA$^{-1}$. (b) Top: valence band of MLG and its characteristic cross sections. The red dotted and yellow and green dashed lines show energy contours for cuts indicated by gray planes. Bottom: simulated MLG ARPES constant-energy maps at energies of the cuts above. The saddle points in MLG dispersion are located at $\vect{M}$.}
	     \label{fig:topology}
\end{figure*}

The importance of interlayer coupling and mSL effects in our structures is most strikingly captured by the constant-energy maps of ARPES intensity at energies $\sim 2.5$ eV below the Dirac points of the layers, shown in Fig.~\ref{fig:topology}(a) for tBLG-C (experimental data in colours, simulation in black and white; we present constant-energy maps for tBLG-A and tBLG-B in SI). For comparison, evolution of the constant-energy line of monolayer graphene (MLG) is shown in Fig.~\ref{fig:topology}(b). In undoped MLG, the constant-energy surface at the Fermi energy, $\epsilon=0$, consists of points, known as Dirac points, located at the corners of the hexagonal Brillouin zone (BZ) and marked as $\vect{K}_{1}$ and $\vect{K}_{1}^{'}$ in the figure. For decreasing energy of the cut, each of the Dirac points gives rise to a closed contour, indicated with red dotted lines in the first plane cutting through the MLG dispersion in Fig.~\ref{fig:topology}(b). Overall, two closed contours can be built from the pieces within the BZ, as seen in inset (i) below the MLG dispersion in which the contours and BZ shown in blue solid line are overlaid on the simulated ARPES intensity map for the same energy (note the crescent-like patterns of intensity around each valley, reflecting the topological nature of the Dirac points \cite{mucha-kruczynski_prb_2008}). The contours grow away from the Dirac points and connect at the $\vect{M}$ points at the energy $\epsilon=\epsilon_{\vect{M}}$ corresponding to the position of cut (ii) in Fig.~\ref{fig:topology}(b). For energies $\epsilon<\epsilon_{\vect{M}}$, cut (iii), only one closed contour is present inside the BZ. 

It is clear from the ARPES spectra in Fig.~\ref{fig:topology}(a) that topology of large-angle tBLG bands is different. For energies $0>\epsilon\gtrsim -1.5$ eV, panel (i), ARPES maps show twelve crescent-like shapes indicating twice the number of Dirac points, in agreement with the presence of two graphene layers. The six less intense features come from the bottom graphene layer, signal from which is attenuated due to the electron escape depth effect. At the energy $\epsilon\approx -2.0$ eV, the crescent shapes connect with each other and states belonging to different layers hybridize. This leads to the formation of one contour encircling the $\Gamma$ point, similarly to MLG at $\epsilon<\epsilon_{\vect{M}}$, as well as, at energy $\epsilon\approx -2.4$ eV, panel (ii), to additional intense features indicated with black arrows. These intense features evolve into new crescent shapes as shown in panel (iii), $\epsilon=-2.78$ eV, and the intensity patterns look strikingly similar to those in panels (a)(i) and (b)(i), suggesting the presence of secondary Dirac points akin to those detected in small-angle tBLG \cite{berdyugin_sciadv_2020} or graphene aligned to underlying $h$-BN \cite{wallbank_prb_2013, wang_naturephys_2016, mucha-kruczynski_prb_2016}. The crescent-like patterns merge together at $\epsilon\approx -3.1$ eV so that for $\epsilon\lesssim\epsilon_{\vect{M}}$, panel (iv), the constant-energy maps contain two concentric contours. These are a consequence of hybridization of the approximately circular and degenerate bottoms of the $\pi$-bands of the two layers due to interlayer coupling with the states shifted to higher (lower) energies giving rise to the inside (outside) contour. 

\begin{figure*}[!t]
	     \centering
	     \includegraphics[width=1.00\linewidth]{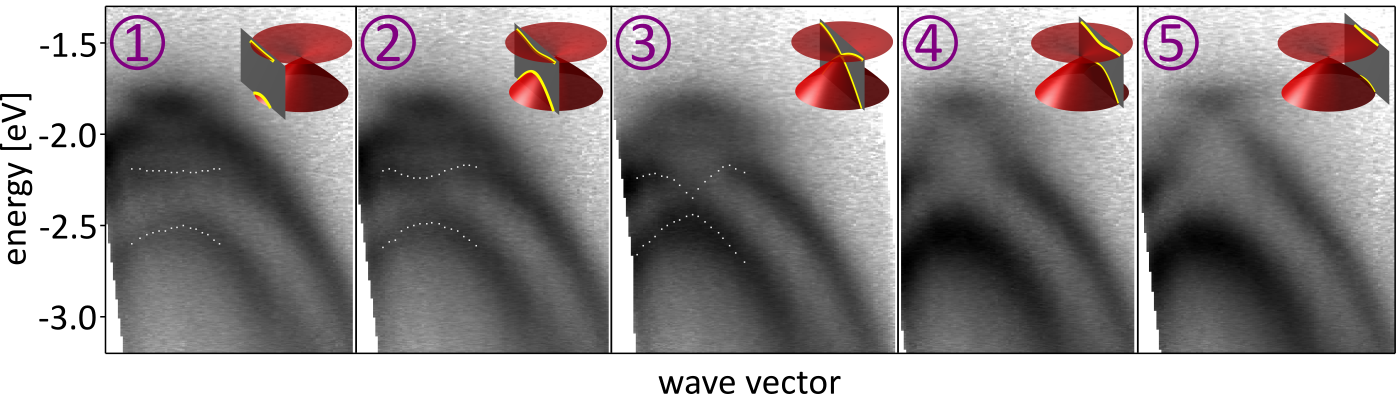}
	     \caption{Secondary Dirac point in large-angle tBLG. Photoemission intensity along wave vector cuts in the vicinity of one of the secondary Dirac points discussed in the main text, as shown with black dashed lines and numbered in Fig.~\ref{fig:topology}(a)(ii). The insets show schematically the shape of the two bands at the energy $\sim-2.5$ eV, with the gray planes indicating the location of the cut with respect to the sDP and the yellow lines highlighing the band cross-section for a given cut. The white dots in cuts 1-3 mark positions of Gaussian peaks fitted to the data to establish the band dispersion.}
	     \label{fig:sDP}
\end{figure*}

We investigate the secondary Dirac points from Fig.~\ref{fig:topology}(a)(ii) in more detail by studying cuts marked 1-5 in that panel and show their photoemission maps in Fig.~\ref{fig:sDP}. For cuts 1-3, we fitted the positions of two bands around the energy $\sim -2.5$ eV with Gaussians (see SI for a description of the procedure), with their peaks as a function of wave vector marked with white dots. Our cuts suggest band structure feature containing a Dirac point as shown in the insets of each panel, where the gray planes indicate the location of the cut and the yellow lines highlight the band cut giving rise to the corresponding ARPES intensity. Our photoemission data cannot exclude the possibility that the secondary Dirac point is gapped; if so, the gap is smaller than $\sim 0.2$ eV (limit imposed by our energy resolution and precision of the fitting procedure). Finally, while the symmetry of the constant-energy maps in Fig.~\ref{fig:topology} implies that the band structure in cuts 1 and 2 is the same as in cuts 4 and 5, we do not see the band above the secondary Dirac point in the latter -- this is because intensity from this part of the dispersion is affected by the Berry phase interference effects \cite{mucha-kruczynski_prb_2008} responsible for crescent-like intensity patterns from otherwise circular contours in the vicinity of Dirac points in the maps in Fig.~\ref{fig:topology}. In the SI, we show additional cuts in the vicinity of the new Dirac point in the direction roughly perpendicular to cuts in Fig.~\ref{fig:sDP}. 

\begin{figure*}[!t]
	     \centering
	     \includegraphics[width=1.00\linewidth]{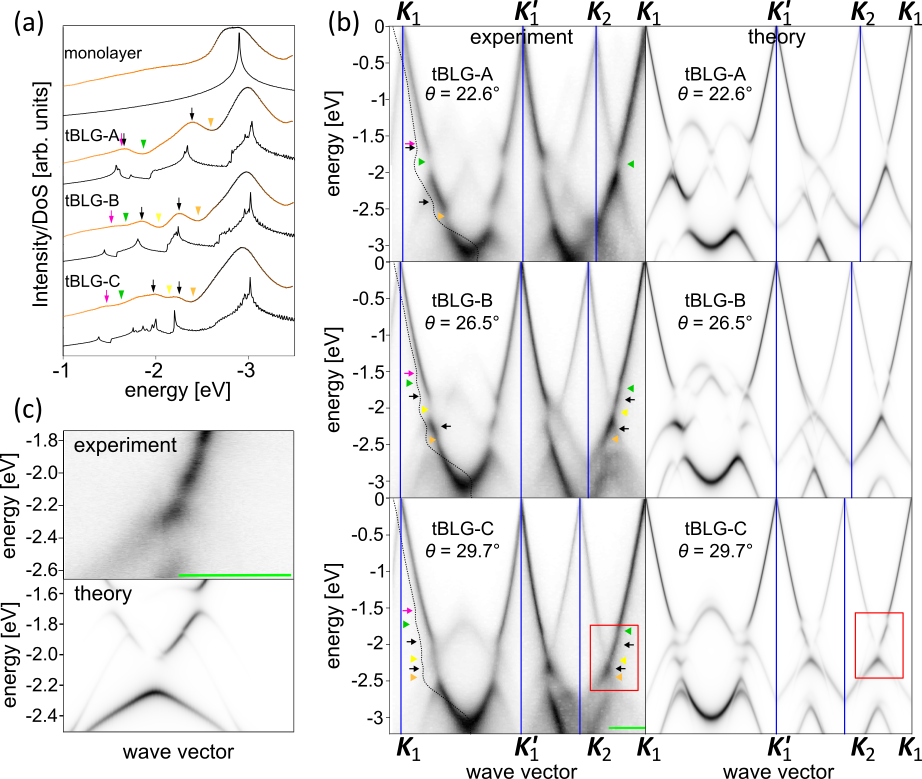}
	     \caption{Minigaps in large-angle tBLG. (a) Energy distribution curves and simulated DoS for MLG and tBLG. Arrows and triangles indicate positions of vHs and minigaps with colours differentiating between the origin of the features as discussed in the text. (b) ARPES intensity along $k$-space path shown with green dashed line in Fig.~\ref{fig:topology}(a), together with the corresponding theoretical simulation (right). The dotted lines are energy distribution curves from (a) with coloured markers indicating the same features. (c) Closeup of the area marked with the red rectangle in (b). The green scale bars in (b) and (c) correspond to 0.5 \AA$^{-1}$.}
	     \label{fig:dos_and_gaps}
\end{figure*}

Changes in the topology of the constant-energy contours like these presented in Fig.~\ref{fig:topology}(a) are reflected by discontinuities in the electronic density of states (DoS): merging of two contours involves a saddle point and generates a van Hove singularity peak (vHs) while appearance of a new one generates a step due to a contribution from a new band. With this in mind, we study the photoemission energy distribution curves obtained by integration of the photocurrent across $k$-space. In Fig.~\ref{fig:dos_and_gaps}(a), we compare the results for all tBLG samples as well as a reference monolayer region of one of the samples and DoS calculated using the continuum model \cite{santos_prl_2007, bistritzer_pnas_2011, koshino_njp_2015} (see SI for description of the theoretical model). The MLG DoS displays a single peak, in the ARPES data reflected as a broad "bump", which corresponds to the saddle points at $\vect{M}$. The large width of this feature for MLG as compared to the theoretical DoS is due to the contribution from the valence band of $h$-BN with its band edge $\sim2.7$ eV below the graphene Dirac points responsible for the left side of the peak (while the $h$-BN signal is strongly attenuated for tBLG, this is less so for the MLG with only one graphene layer on top of the substrate). A similar feature originating from the $\vect{M}$ saddle points is also present at slightly shifted positions in all the tBLG DoS. However, the tBLG curves contain additional features indicating the presence of several vHs singularities and suggesting a more complicated band structure evolution than evident from the constant-energy maps. Positions of these features are well correlated with sharp peaks in the theoretical DoS below each experimental plot -- we highlight with arrows the maxima and with triangles the minima of photocurrent that are of special interest below. 

The top curve in Fig.~\ref{fig:dos_and_gaps}(a) was obtained by moving the nano-ARPES spot off the region where the two layers overlap. This provides a direct comparison between monolayer and twisted bilayer and suggests that the changes in the photocurrent measured from tBLG areas are purely due to the interlayer interaction. In van der Waals heterostructures with twisted interfaces, two mechanisms are known to induce DoS peaks: (i) direct hybridization of states from different layers \cite{li_naturephys_2010} and (ii) coupling between states backfolded by the mSL \cite{garcia-ruiz_arxiv_2020}. Both lead to opening of gaps in the electronic spectrum as a consequence of coupling between electronic states, accompanied by the appearance of saddle points in the dispersion which in turn are responsible for the DoS peaks. Therefore, to understand the energy distribution curves in Fig.~\ref{fig:dos_and_gaps}(a), we look for signs of minigap formation by investigating photoemission spectra along the $k$-space paths connecting the valleys $\vect{K}_{1}$, $\vect{K}^{'}_{1}$ and $\vect{K}_{2}$ as shown in Fig.~\ref{fig:topology}(a). We present these cuts in Fig.~\ref{fig:dos_and_gaps}(b), together with simulations produced using a model established for ARPES studies of graphene on $h$-BN \cite{mucha-kruczynski_prb_2016} and applied to graphene stacks \cite{thompson_naturecomms_2020, zhu_arxiv_2020} (see SI for details). The theoretical model captures all the qualitative features of the experimental data. Moreover, DoS minima in (a) coincide with the positions of the minigaps in (b). Because opening of minigaps in the electronic spectrum of two-dimensional materials must be accompanied by generation of saddle points, we identify the DoS maxima with a vHs in the vicinity of each minigap. For devices tBLG-C and tBLG-B, we can resolve at least three minigaps, as shown in more detail for the former in panel (c) which presents a separate measurement of the region indicated by the red rectangle in (b). This implies observation of four minibands, a testament of the outstanding quality of our samples. 

\begin{figure*}[t]
	     \centering
	     \includegraphics[scale=1.00]{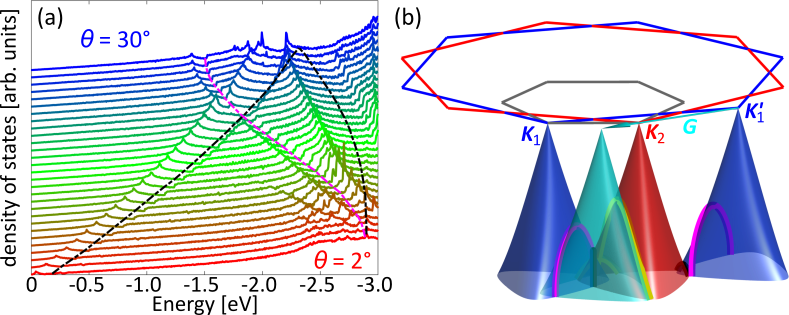}
	     \caption{Moir\'{e}-induced scattering in large-angle tBLG. (a) Evolution of the tBLG DoS for $\theta=2^{\circ}$ (red) to $\theta=30^{\circ}$ (blue), in steps of $1^{\circ}$ (curves shifted vertically for clarity). The dashed lines are guides for the eye indicating, for given $\theta$, highest energies of the crossings marked with the corresponding colour in (b). (b) Hierarchy of crossings in tBLG with $\theta>\theta_{c}$. The blue and red hexagons are the BZ of the top and bottom graphene layer; their valence band structures in the vicinity of $\vect{K}_{1}$, $\vect{K}^{'}_{1}$ and $\vect{K}_{2}$ are shown with blue and red surfaces, respectively. The cyan cone depicts the $\vect{K}^{'}_{1}$ states shifted by a moir\'{e} reciprocal vector $\vect{G}$ indicated with the cyan arrow (the moir\'{e} BZ is shown in gray). Crossings between MLG dispersions are highlighted in black (between two MLG dispersions twisted by $\theta$), magenta ($\vect{K}_{1}$ cone and $\vect{K}_{1}^{'}$ translated by $\vect{G}$; $\vect{K}_{1}^{'}$ cone and $\vect{K}_{1}$ translated by $-\vect{G}$) and yellow ($\vect{K}_{2}$ cone of bottom MLG and $\vect{K}_{1}^{'}$ translated by $\vect{G}$).}
	     \label{fig:umklapp}
\end{figure*}

To discover the origin of the observed minigaps and vHs, we study the evolution of DoS calculated for twists $2^{\circ}\leq\theta\leq30^{\circ}$, in steps of $1^{\circ}$, shown in Fig.~\ref{fig:umklapp}(a) (curves have been shifted vertically for clarity). In the absence of interlayer coupling, two MLG dispersions rotated with respect to each other by $\theta$ must intersect and we mark such crossings in black in Fig.~\ref{fig:umklapp}(b) where we show conical valence band dispersions of the top (blue) and bottom (red) layers for $\theta=26.5^{\circ}$. The neighbouring Dirac points are separated by a distance $\tfrac{8\pi}{3a}\sin\tfrac{\theta}{2}$ \cite{santos_prl_2007} ($\vect{K}_{1}$ and $\vect{K}_{2}$ as marked in Fig.~\ref{fig:topology}), where $a$ is the graphene lattice constant, or $\tfrac{8\pi}{3a}\sin\tfrac{\pi/3-\theta}{2}$ ($\vect{K}_{2}$ and $\vect{K}^{'}_{1}$). The highest energies of crossings occur midway between every pair of Dirac points and the corresponding energies as a function of $\theta$ are indicated with the black dashed lines on top of the DoS curves in (a). Interlayer coupling hybridizes the degenerate states at the crossings, turning them into anti-crossings accompanied by a saddle point between the Dirac points and above the gap (note that the saddle point is shifted off the line connecting the Dirac points \cite{garcia-ruiz_arxiv_2020}) and a quasi-quadratic edge of the next miniband below. The corresponding DoS features, peak at higher energies due to the saddle point and a step at lower energies due to the band edge, can be seen in the vicinity of both dashed black lines in the DoS curves in (a) (the hybridization minigap does not open a global band gap as other parts of the electronic dispersion overlap with it so that the electronic density of states does not go down to zero \cite{brihuega_prl_2012, li_naturephys_2010, bistritzer_pnas_2011}). At small twist angles, the feature closest to the Dirac points is due to mixing of states between pairs of Dirac cones closest to each other and has been studied using scanning tunnelling spectroscopy \cite{li_naturephys_2010, brihuega_prl_2012}, ARPES \cite{ohta_prl_2012, yin_naturecomms_2016} and magnetic focusing \cite{berdyugin_sciadv_2020}. At larger twists, separations between all pairs of neighbouring Dirac cones become comparable, driving the associated vHs into the energy range $\sim 2$ eV from the Dirac points. Guided by the approximate positions of minigaps indicated by the black dashed lines in Fig.~\ref{fig:umklapp}(a), we ascribe the ARPES features marked with black arrows in Fig.~\ref{fig:dos_and_gaps}(a) to vHs formed above direct-hybridization gaps while the gaps themselves correspond to features indicated with the yellow and orange triangles.

Interestingly, for tBLG-B and tBLG-C, the ARPES features marked in Fig.~\ref{fig:dos_and_gaps}(a) with magenta arrows and green triangles cannot be explained by mixing of degenerate electronic states of the two layers by interlayer coupling. Instead, they evidence scattering of electrons by the mSL. In Fig.~\ref{fig:umklapp}(b), we show in gray the moir\'{e} BZ in relation to the BZ of the graphene layers (red and blue for bottom and top, respectively). The primitive reciprocal vectors of the mSL correspond to the shortest vectors produced by subtraction of the reciprocal vectors of the two crystals \cite{santos_prl_2007, wallbank_annphys_2015}, with one such vector, $\vect{G}$, portrayed by the cyan arrow. Scattering of electrons from the valley $\vect{K}^{'}_{1}$ of the top layer by that moir\'{e} reciprocal vector can be schematically depicted by translating the whole cone, producing the cyan surface which intersects with conical dispersion surfaces of the top layer around $\vect{K}_{1}$. We mark this intersection with a magenta line on the $\vect{K}_{1}$ cone. We also mark in the same colour on $\vect{K}_{1}^{'}$ cone the equivalent intersection of $\vect{K}_{1}^{'}$ dispersion with $\vect{K}_{1}$ translated by $-\vect{G}$. The highest energy of these crossings, midway between $\vect{K}_{1}$ and translated $\vect{K}^{'}_{1}$ (or between $\vect{K}_{1}^{'}$ and translated $\vect{K}_{1}$), is indicated as a function of $\theta$ with the dashed magenta line in Fig.~\ref{fig:umklapp}(a) and provides an estimate for the position of a vHs formed above a minigap opened due to the moir\'{e}-induced intervalley interaction of $\vect{K}^{'}_{1}$ electrons with those in $\vect{K}_{1}$. For small twists, the primitive reciprocal vectors of mSL are short and $\vect{K}^{'}_{1}$ replica intersects the original dispersion of the top layer far below the Dirac points. The energy of the intersection increases with increasing twist angle as the moir\'{e} reciprocal vector scatters $\vect{K}^{'}_{1}$ electrons closer to $\vect{K}_{1}$. At $\theta=\theta_{c}=\mathrm{arccos}\tfrac{13}{14}\approx 21.8^{\circ}$, the distance between $\vect{K}_{1}$ and $\vect{K}^{'}_{1}$ replica is the same as between $\vect{K}_{1}$ and $\vect{K}_{2}$ so that the highest energies of the corresponding intersections are at similar energies (the energies are not identical because of the trigonal warping of the cone-like dispersions). This means that the related minigaps and vHs should also overlap as is indeed the case for DoS of sample tBLG-A with $\theta=22.6^{\circ}$ in Fig.~\ref{fig:dos_and_gaps}(a). For larger twist angles, scattering on the moir\'{e} potential brings the $\vect{K}^{'}_{1}$ states close enough to $\vect{K}_{1}$ so that it is this process, rather than direct hybridization of $\vect{K}_{1}$ and $\vect{K}_{2}$ cones, that is responsible for the ARPES features closest to the Dirac points in tBLG-B and tBLG-C: minigaps indicated with green triangles and vHs marked with magenta arrows in Fig.~\ref{fig:dos_and_gaps}. 

\begin{figure*}[t]
	     \centering
	     \includegraphics[width=1.00\linewidth]{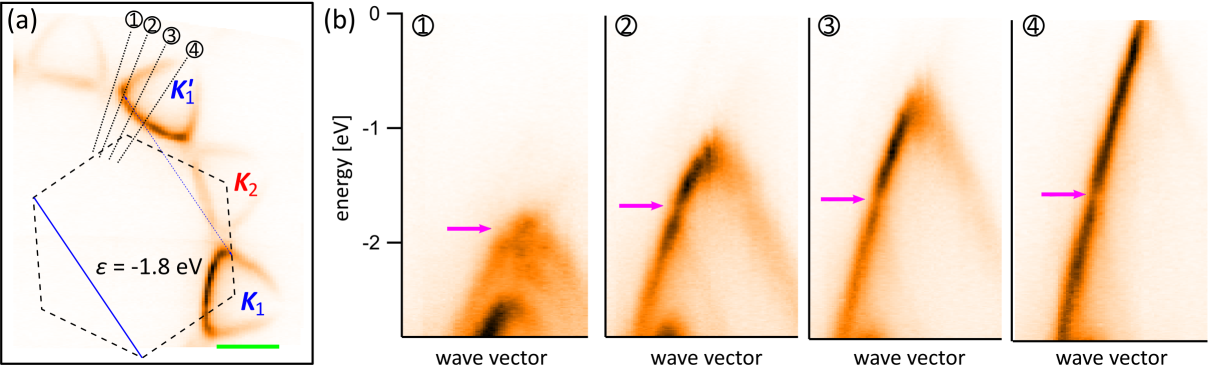}
	     \caption{Tracking mSL minigaps in electronic dispersion. (a) Constant-energy map for tBLG-C for energy $\epsilon=-1.8$ eV showing coupling between states related by a moir\'{e} reciprocal vector (thin blue line). The moir\'{e} BZ is drawn in black dashed lines with the same moir\'{e} reciprocal vector presented in blue for comparison. The green scale bar corresponds to 0.5 \AA$^{-1}$. (b) Photointensity measured along the cuts 1-4 marked in panel (a). The magenta arrows indicate, for each cut, the position of the minigap formed due to the moir\'{e}-induced coupling between states in the $\vect{K}_{1}$ and $\vect{K}_{1}^{'}$ valleys of the top graphene layer.}
	     \label{fig:parabolas}
\end{figure*}

Further confirmation that moir\'{e}-induced scattering is responsible for some of the minigaps and vHs we observe can be provided by explicitly connecting affected states with mSL reciprocal vectors. In the constant-energy map in Fig.~\ref{fig:parabolas}(a), corresponding to the energy marked by the green triangle for tBLG-C in Fig.~\ref{fig:dos_and_gaps}(b), $\epsilon=-1.8$ eV, we connect positions of the minigaps around $\vect{K}_{1}$ and $\vect{K}^{'}_{1}$ with the moir\'{e} reciprocal vector $\vect{G}$ (thin blue line; see SI for procedure used to determine the moir\'{e} BZ). Moreover, in panel (b) we show photoemission measured along the cuts 1-4 as numbered and marked in (a). Using these cuts, we can trace the crossing of $\vect{K}_{1}^{'}$ cone with the $\vect{K}_{1}$ one translated by $-\vect{G}$ and the resulting minigap, indicated with the magenta arrow for each cut, effectively following the magenta line on the $\vect{K}_{1}^{'}$ cone in Fig.~\ref{fig:umklapp}(b). Note that we do not observe any minigaps (or features in the experimental and theoretical DoS) due to the hypothetical crossings between $\vect{K}^{'}_{1}$ states scattered by moir\'{e} superlattice and bottom layer dispersion around $\vect{K}_{2}$ [yellow line in Fig.~\ref{fig:umklapp}(b)]. This is because such process is higher order in the mSL perturbation (it involves additional interlayer tunnelling). Finally, scattering of bottom layer electrons on the potential of the top layer [moir\'{e}-induced coupling between $\vect{K}_{2}$ and $\vect{K}^{'}_{2}$; not shown in Fig.~\ref{fig:umklapp}(b)] is difficult to observe because of the additional attenuation of the signal from the bottom layer. 

With regards to the magnitudes of the minigaps, the largest direct hybridization gap we observe is the one for tBLG-C shown in Fig.~\ref{fig:dos_and_gaps}(c), $\Delta_{\mathrm{direct}}\sim 0.25$ eV. For intermediate twist angles, $1^{\circ}\ll\theta\ll 30^{\circ}$, an estimate of this gap can be obtained {\it via} degenerate perturbation theory for two states coupled by $t\approx 0.11$ eV \cite{santos_prl_2007, bistritzer_pnas_2011}, yielding $\Delta_{\mathrm{direct}}=2t\approx 0.22$ eV. At small angles, the moir\'{e} wave vector rapidly decreases with decreasing twist angle so that mSL couples states on Dirac cones with energy separation $\epsilon\ll t$. Level repulsion between these densely packed states leads to miniband separations decreasing with $\theta$ and flat bands in the extreme limit of the magic angle \cite{bistritzer_pnas_2011}. In turn, at large angles, as discussed earlier the moir\'{e} vectors are sufficiently long to couple the two degenerate states to states from other valleys, some with energies within $\sim t$ from the band crossing. Such states lead to a slight increase of the hybridization minigap with angle and a complicated band structure in its vicinity with several additional (moir\'{e}-induced) minigaps as observed for tBLG-B and tBLG-C (in the limit of $\theta=30^{\circ}$, moir\'{e} couples 12 equi-energetic states from both graphene layers \cite{moon_prb_2019}). To estimate the size of the minigap opened due to moir\'{e}-induced scattering, $\Delta_{\mathrm{moir\acute{e}}}$, one must consider at least three states: two degenerate states on the crossing of the $\vect{K}_{1}$ and $\vect{K}^{'}_{1}+\vect{G}$ cones [magenta line at the crossing of the blue and cyan cones in Fig.~\ref{fig:umklapp}(b)] and an electronic state of the bottom layer at the same wave vector and at energy ${\Delta\epsilon}\sim 1$ eV away. The first two states are not directly coupled to each other but only to the third one through the interlayer coupling $t$, so that $\Delta_{\mathrm{moir\acute{e}}}\sim\tfrac{2t^{2}}{\Delta\epsilon}\approx0.02$ eV. Note that such a minimal three-level model underestimates the moir\'{e}-induced gaps we observe. We discuss our estimates for $\Delta_{\mathrm{direct}}$ and $\Delta_{\mathrm{moir\acute{e}}}$ in more detail in the SI.

We have checked that the suppression of photocurrent we identify with spectral minigaps cannot be ascribed to photoemission final state effects \cite{kurtz_jpcm_2007} which include dependence of the intensity on photon energy as well as polarization \cite{gierz_nanolett_2012, liu_prl_2011, gierz_prb_2011, hwang_prb_2011, damascelli_rmp_2003}. In the SI, we show single-particle spectral weight of the electronic wave function for the wave vector and energy range as used for the ARPES spectra in Fig.~\ref{fig:dos_and_gaps}(c). This spectral weight contains all of the minigaps discussed here which demonstrates that these are true spectral features and do not arise as a result of suppression of photointensity due to the Berry phase or final state effects. Moreover, we have performed measurements using photons both with energies 27 and 74 eV and observed little change in the spectra (a comparison of a cut along the $\vect{K}_{1}-\vect{K}_{1}^{'}$ direction for sample tBLG-C measured at both photon energies is shown in the SI). We use linearly-polarized light and our geometry is such that when measuring along the $\vect{\Gamma}-\vect{K}_{1}$ direction the detector is in the plane of incidence and the incident light is $p$-polarized. This determines directions in the reciprocal space along which the photointensity  is suppressed due to the Berry phase associated with the BZ corners \cite{mucha-kruczynski_prb_2008, gierz_nanolett_2012, liu_prl_2011, gierz_prb_2011, hwang_prb_2011} [in the valence band, starting from a BZ corner in the direction away from $\Gamma$, as evident in the maps in Fig.~\ref{fig:topology}(a)(i)] and allows us to confirm that these do not overlap with locations of the minigaps, see for example Fig.~\ref{fig:parabolas}. Our observations also cannot be the result of secondary scattering of photoelectrons as this leads to band replicas but not gap opening \cite{mucha-kruczynski_prb_2016}.

\section{Conclusions}

Our results demonstrate the robustness of the moir\'{e} superlattice picture at large twist angles when the moir\'{e} wavelength $\lambda=a/[2\sin(\tfrac{\theta}{2})]$ is comparable to the graphene lattice constant, $a$, and cannot correspond to a lattice constant of a commensurate superlattice. In large-angle twisted bilayer graphene with $\theta>21.8^{\circ}$, gaps opened by the moir\'{e}, together with the associated van Hove singularities, are the closest to the Dirac points density of states features evidencing interaction of the two graphene layers. The direct hybridization minigap is located deeper in the valence band and is the largest for tBLG-C with the twist close to $30^{\circ}$, $\Delta_{\mathrm{direct}}\sim 0.25$ eV. The interlayer coupling also modifies the topology of the dispersion at energies in the vicinity of and below the Brillouin zone $\vect{M}$ points where the quasi-conical dispersions merge: we observe secondary Dirac points in the reconstructed spectrum as well as hybridization of the bottom parts of the valence bands. It is worth noting that the LEED spectra shown in SI indicate no strain reconstruction in our graphene crystals. 

While a signature of moir\'{e}-induced scattering was observed previously for a $30^{\circ}$ twisted bilayer graphene with its aperiodic moir\'{e} \cite{yao_pnas_2018}, we show that these processes are not restricted to this special twist angle but rather provide a robust way of coupling electronic states, with the twist angle controlling the affected regions of reciprocal space. The recent work on graphene on InSe \cite{graham_arxiv_2020} suggests that moir\'{e}-induced scattering is not limited to twisted bilayer graphene. 

Finally, it has been shown that it is possible to dope MLG sufficiently to move the chemical potential to the $\vect{M}$ point van Hove singularity \cite{mcchesney_prl_2010, link_prb_2019, rosenzweig_prb_2019, rosenzweig_prl_2020} and so it might be feasible to explore large-angle tBLG in a similar regime. Interestingly, a superconducting instability was predicted for the MLG doped to the vHs \cite{mcchesney_prl_2010} but magnetic ordering for tBLG doped to the Dirac cone anti-crossing \cite{gonzalez_prb_2013} (situation not equivalent to magic-angle tBLG in which states coupled by moir\'{e} reciprocal vectors contribute significantly to the flat bands \cite{bistritzer_pnas_2011}), with recent experimental studies in agreement with the latter \cite{liu_prb_2019}. This suggests large-angle tBLG as a platform in which the interaction effects at vHs of different origin (in-plane nearest-neighbour coupling, interlayer Dirac cone anti-crossing, moir\'{e}-induced intralayer intervalley coupling) and competition between them could be explored. 

\section{Methods}

First, laterally large ($>100\,\mu$m) and thin ($<100$ nm) $h$-BN was mechanically exfoliated onto a Ti/Pt (2/10nm) coated highly $n$-doped silicon wafer. Monolayer graphene was then transferred onto the $h$-BN using the poly(methyl methacrylate) (PMMA) dry peel stamp transfer technique \cite{ponomarenko_natphys_2011, frisenda_csr_2018}. To note, a few-layer graphene (connect to the monolayer) overlapped the edge of the $h$-BN to form a ground to the highly conductive Ti/Pt/Si substrate. A second graphene flake was then deterministically transferred onto the stack to create the tBLG. The stack was then annealed at $300\,^{\circ}$C for 3 hours to allow contamination trapped between flakes to agglomerate through the self-cleaning mechanism \cite{haigh_natmat_2012}. The LEEM, LEED and ARPES measurements were performed at the Elettra Synchrotron \cite{mentes_bjn_2014, dudin_jsr_2010}. All ARPES spectra in the main text were obtained using photons with energy of 74 eV, except Fig.~\ref{fig:dos_and_gaps}(c) which has been obtained with 27 eV photons. 

To simulate the ARPES spectra, we used the tight-binding model to describe each of the graphene layers coupled with a continuum description of the interlayer interaction \cite{santos_prl_2007, bistritzer_pnas_2011, koshino_njp_2015}. The layers are considered  rigid (which is a good approximation at large angles for which variation of the interlayer distance across the moir\'{e} unit cell decreases to $\simeq 0.01$ \AA \cite{gargiulo_2dmaterials_2018}) and the interlayer coupling is taken into account {\it via} the Fourier transform of the Slater-Koster-like \cite{slater_physrev_1954} hopping between $p$-orbitals. Values of the parameters in our model are based on those used previously in the literature and shown to be applicable to a large range of twist angles \cite{koshino_njp_2015, thompson_naturecomms_2020} as well as the fit to the experimental ARPES data. The detailed discussion of the procedure used to produce photocurrent intensity is presented in SI.

\section{Author Contributions}

M.H.~and R.G.~fabricated the samples. A.G., V.K., F.G., T.O.M., A.L.~and A.B.~performed the LEED, LEEM and ARPES measurements. M.M.-K.~built the theoretical model and simulated the ARPES spectra. A.B.~and M.M.-K.~analysed the ARPES data and wrote the manuscript with input from all the authors.

\section{Acknowledgements}

The authors thank N.~Wilson and V.~Fal'ko for insightful discussions. We acknowledge support from the Engineering and Physical Sciences Research Council (EPSRC) grant EP/V007033/1, the European Graphene Flagship Project, European Quantum Technology Flagship Project 2D-SIPC (820378) and the Royal Society. M.M.-K.~acknowledges support by the University of Bath International Research Funding Scheme.

\section{Supporting Information}

The Supporting Information is available free of charge at \href{https://pubs.acs.org/doi/10.1021/acsnano.1c06439}{https://pubs.acs.org/doi/10.1021/acsnano.1c06439}.

Further details of the LEEM, LEED and ARPES measurements, determination of the twist angle, constant-energy ARPES maps for tBLG-A and tBLG-B, additional ARPES spectra for cuts across the secondary Dirac point and comparison of spectra taken with photons of different energy; description of the theoretical model used to obtain tBLG bands and ARPES spectra; effective model used to estimate magnitudes of the observed minigaps.

%

\end{document}


\beginsupplement
	
\maketitle

\newpage
\tableofcontents

\newpage

%

\begin{figure*}[!b]
    \centering
    \includegraphics[width = 1.0\linewidth]{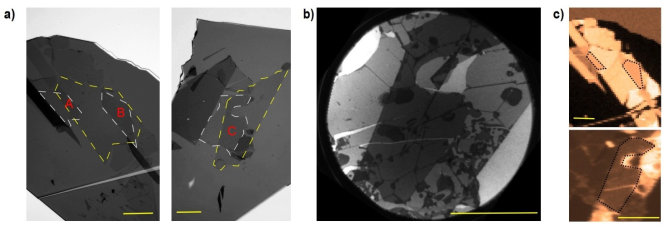}
    \caption{Selected optical (a), LEEM (b) and SPEM (c) images of the samples. In the optical micrographs, bottom graphene layers are delineated in yellow and the top graphene layers are in white. The overlap of the two layers determines tBLG-A, B and C devices as marked in the figure and delineated by black dotted polygons in (c). The scale bar in each image is 20 $\mu$m.}
    \label{fig:S1}
\end{figure*}

\section*{Experimental measurements}

Before the measurements the samples were transferred in ultra-high vacuum and annealed for several hours at $400$-$450^{\circ}$C to remove surface contamination. The measurements were performed in ultra-high vacuum of less than $2\times10^{-10}$ mbar with the sample at room temperature for LEEM-LEED and at 95 K for ARPES.

While annealing can in some cases reorient graphene on $h$-BN \cite{wang_prl_2016}, we have no reason to think this has affected our samples in between LEED and ARPES measurements. The first annealing process was performed during fabrication and adjustments of the relative positions of the flakes, if any, happened at that point. We have performed ARPES and LEED measurements several times on the same locations, with the samples annealed before every measurement and did not observe any changes in the measured twist angles or ARPES spectra.

\subsection*{LEEM and $\mu$-LEED}

The sample morphology and crystal structure were studied using the SPELEEM microscope at the Nanospectroscopy beamline at Elettra \cite{mentes_bjn_2014}. In particular, low energy electron microscopy \cite{bauer_rps_1994} was used to image the graphene and $h$-BN crystal grains, visualizing boundaries and defects [Fig.~\ref{fig:S1}(b)]. In this manner, defect-free areas of homogeneous quality were selected for further LEED analysis. The LEED patterns were measured {\it in situ}, operating the microscope in diffraction mode. A suitable illumination aperture was used to limit the e-beam footprint on the sample, allowing us to probe a circular area of about 1 micron in diameter.

\begin{figure*}[t]
    \centering
    \includegraphics[width = 1.0\linewidth]{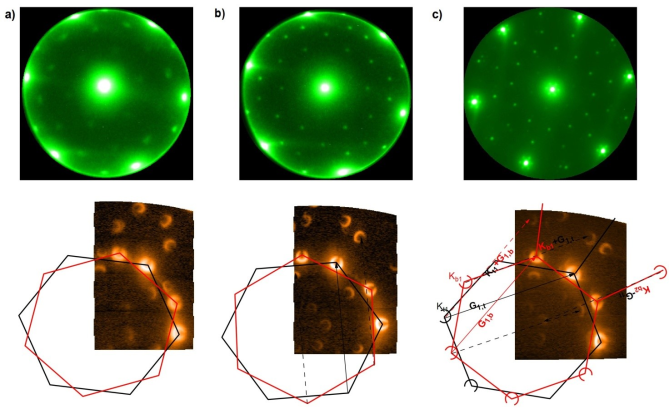}
    \caption{$\mu$-LEED patterns with 45 eV electrons (top) and $\mu$-ARPES constant energy maps at 0.8 eV (bottom) below DP for tBLG-A, B and C [panels (a), (b) and (c), respectively]. All intensities are plotted in logarithmic scale. BZs of top (bottom) layers are shown schematically as black (red) hexagons.}
    \label{fig:S2}
\end{figure*}

Careful analysis of the LEED patterns in Fig.~\ref{fig:S2} provides some information about the relative alignment between the $h$-BN and the bottom graphene layer. For example, for tBLG-C, a ring of six spots surrounding the brighter features of the bottom layer allows to obtain $\theta_{h-\mathrm{BN}}$.

\subsection*{$\mu$-ARPES}

In order to take angle resolved photoemission spectra from micrometer size areas on the samples ($\mu$-ARPES), the synchrotron radiation light was focused to a 600 nm spot using Schwarzschild objectives with multilayer coated spherical mirrors optimized for photons of 27 and 74 eV. The photoelectron angle and energy distribution maps were obtained with a movable hemispherical electron energy analyzer \cite{dudin_jsr_2010}. To locate the region of interest for ARPES, scanning photoemission images (SPEM) were taken [Fig.~\ref{fig:S1}(c)] with the angle and energy of the electron energy analyzer set to count selected graphene $\pi$-band(s): on the bottom image, the bottom layer graphene band distribution is shown; on top image the bands of both layers contribute to the intensity. Confronting the contrast variations in SPEM with optical and LEEM images allows to identify unambiguously the overlap of top and bottom graphene layers. The same points for both $\mu$-LEED and $\mu$-ARPES were selected to avoid possible differences in twist angle due to wrinkles and grains in exfoliated graphene flakes. 

\subsection*{Twist angle measurements and Dirac cone replicas in LEED and ARPES maps}

The twist angle $\theta$ can be selected during the fabrication process by aligning the edges of graphene flakes under the optical microscope. This leads to a potential error of $30^{\circ}$, depending on whether the edges of top and bottom layers are same or different of the two possible types, armchair or zigzag. Moreover, sub-degree resolution is necessary to compare theoretical calculations and experimental measurements reliably. For these reasons, $\mu$-LEED patterns were used to determine $\theta$ for each sample. For large twist angles, this procedure, performed in the reciprocal space, is more precise than measuring the real space moir\'{e} periodicity with the scanning probe techniques (widely used for small $\theta$) because real space moir\'{e} unit vectors are small, {\it i.e.} comparable with the graphene lattice constant.

The twist angles are calculated from LEED patterns shown in Fig.~\ref{fig:S2} by identifying {\it via} Gaussian fitting the coordinates of two hexagons formed by six main spots from each layer and then finding the rotation angle at which the sum of deviations of distances between pairs of reflections of different layers is minimal. The sum of clockwise and counter-clockwise twist angles found with this procedure was $60^{\circ}$ within less than $0.1^{\circ}$ for all three devices, which we therefore consider the accuracy of our twist angle measurement. 

The maximum rotational disorder within the area of the $\mu$-LEED spot can be evaluated by comparing the (Gaussian) width of the zeroth order diffraction spot (normal electron reflection not affected by rotational misalignment) to the width of 1st order peaks. We find the former to be 0.0166 {\AA}$^{-1}$ while the latter is 0.0181 {\AA}$^{-1}$ and 0.0173 {\AA}$^{-1}$ for the top and bottom graphene layers, respectively. Assuming that the change in width is exclusively due to rotational disorder, its Gaussian distribution would have a width of 0.007 {\AA}$^{-1}$ and 0.0049 {\AA}$^{-1}$ for the top and bottom layers, respectively. This converted into to the angle gives $\Delta\theta_{\mathrm{top}}=0.12^{\circ}$ and $\Delta\theta_{\mathrm{bottom}}=0.08^{\circ}$, comparable with the accuracy of twist angle determination. Overall, the maximum twist disorder cannot be larger than $\Delta\theta=0.2^{\circ}$.

Slightly worse quality of the $\mu$-LEED pattern from tBLG-A is likely because the region used for measurements was near exposed $h$-BN which was charging under the relatively large LEEM electron beam. In comparison, in the case of $\mu$-ARPES, its sub-micron beam was not illuminating $h$-BN during the measurements.

All three devices show secondary LEED reflections and replicas of the characteristic crescent-like patterns in ARPES. In Fig.~\ref{fig:S2}(c), the origin of several $\mu$-ARPES replicas is illustrated: Dirac cones of the bottom layer (red) are scattered by the primitive reciprocal vectors of the top layer (black). The origin of the secondary reflections in LEED is similar. Such features have been initially observed for $\theta=30^{\circ}$ \cite{ahn_science_2018}. Also visible in (c) are additional LEED reflections from the moir\'{e} pattern at the $h$-BN/graphene interface - out of all samples, for tBLG-C the underlying $h$-BN is aligned the closest with the bottom graphene layer ($\theta_{h-\mathrm{BN}}\approx 4^{\circ}$).

\subsection*{Constant-energy ARPES maps for tBLG-A  and tBLG-B}

\begin{figure}[!t]
\centering
\includegraphics[width=0.9\columnwidth]{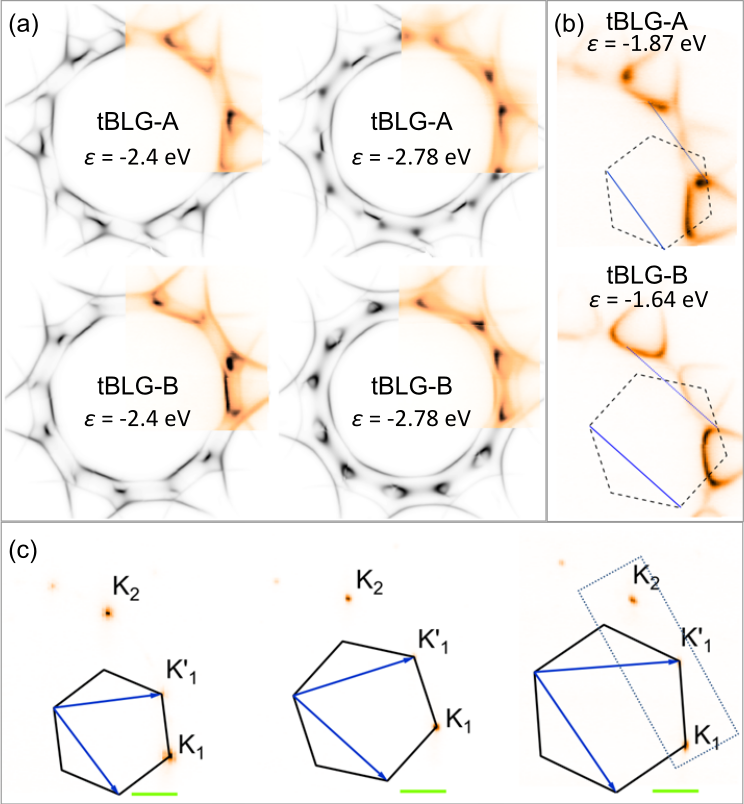}
\caption{(a) Representative ARPES constant-energy maps for devices tBLG-A and B (coloured intensity represents measurement; black and white is simulated). (b) Examples of moir\'{e}-induced scattering between inequivalent valleys of the top graphene layer for tBLG-A and B. The black dashed hexagons denote the corresponding moir\'{e} Brillouin zone and the blue lines indicate one of the moir\'{e} primitive reciprocal vectors. (c) Constant-energy maps at $\epsilon=0$ used to determine the size of the moir\'{e} Brillouin zones for all samples (from left to right, devices A to C). The dashed rectangle indicates the $k$-space area over which the ARPES signal was integrated to obtain the photointensity curves in Fig.~3(a) of the main text. The green scale bars in all panels correspond to 0.5 \AA$^{-1}$.}
\label{fig:S3}
\end{figure}

In Fig.~\ref{fig:S3}(a) we present representative ARPES constant-energy maps for devices tBLG-A and tBLG-B which demonstrate that the evolution of their miniband spectra is qualitatively similar to that for tBLG-C presented in Fig.~1(a) of the main text. In particular, we observe similar crescent-like patterns at energies $\sim 2.5$ eV below the Dirac points suggesting the presence of secondary Dirac points in the electronic spectrum. In contrast to tBLG-C, these appear in two sets of six rather than one set of twelve, with each set forming at a different energy. This is because of the significant deviations of the twist angles from $30^{\circ}$ and its 12-fold symmetry.

In panel (b), we show ARPES constant-energy maps with evidence of moir\'{e}-induced scattering in tBLG-A and tBLG-B. The black dashed hexagons denote the moir\'{e} Brillouin zones and the blue lines indicate the moir\'{e} primitive reciprocal vector which scatters electrons from one valley of the top layer to another (we plot this reciprocal vector twice to show how it fits within the moir\'{e} BZ and what electronic states it connects). The moir\'{e}-induced coupling leads to opening of minigaps which in the maps can be seen as interuption of the crescent-like intensity patterns.  

Finally, in panel (c) we show constant-energy maps at $\epsilon=0$ eV which were used to determine the $\vect{K}_{1}$-$\vect{K}_{2}$ distance and hence the effective moir\'{e} Brillouin zone for each tBLG sample. Once determined, this moir\'{e} BZ was then used for ARPES constant-energy maps at other energies, including the one shown in Fig.~5(a) of the main text. The dotted rectangle in the right-most figure shows $k$-space area which was integrated over to obtain the photointensity curves in Fig.~3(a) of the main text. 

\subsection*{Secondary Dirac point in the miniband spectrum}

In Fig.~2 of the main text, we show cuts through or in the vicinity of the secondary Dirac point we observe in the miniband spectrum of tBLG-C. For the cuts 1-3, we extract the dispersion of the two bands forming this Dirac point (white dots in the corresponding panels in the main text) by fitting the data with Gaussian peaks. We fitted all four bands visible in the images. The widths of the four peaks were similar for the fit of panel 1 and 2 where all four bands have high intensity and can be identified separately. To fit the cut 3 where the intensity of the top band forming the secondary Dirac point considerably drops, we fixed the gaussian widths of the four bands to be as as in panels 1 and 2 to preserve consistency of the fit. Due to the lowered intensity of this band, the error in its position for this cut is $\sim0.2$ eV, which we take as an error in the determination of the gap at the secondary Dirac point.

\begin{figure}[!b]
\centering
\includegraphics[width=1.0\columnwidth]{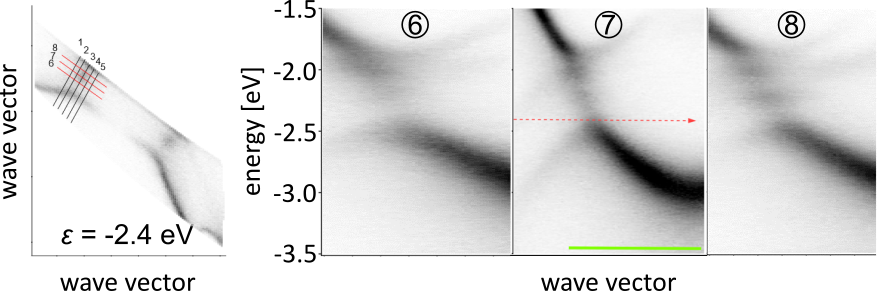}
\caption{Additional cuts through the secondary Dirac point in the miniband spectrum of large-twist tBLG. The left-most panel presents $k$-space directions of cuts 1-5, shown in the main text, and cuts 6-8 shown in the panels to the right as marked on top of each of them. The green scale bar corresponds to $0.5$ \AA$^{-1}$.}
\label{fig:S4}
\end{figure}

In Fig.~\ref{fig:S4}, we present additional $k$-space cuts for tBLG-C in the vicinity of the secondary Dirac point discussed in the main text. These cuts, numbered 6 to 8, are roughly perpendicular to the cuts 1 to 5 shown in Fig.~2 of the main text. Directions of all the cuts are shown together for completeness in the first panel in Fig.~\ref{fig:S4}. The intensity is shown in a linear scale from white to black and the spectra support the notion of a secondary Dirac point in the miniband structure at the energy $\epsilon_{\mathrm{sDP}}\approx-2.4$ eV below the original Dirac points at the graphene Brillouin zone corners. In the cut 7, which passes through one of the spectral features identified as the locations of the secondary Dirac points [see Fig.~1(a)(ii) of the main text], the two bands at the energy $\epsilon_{\mathrm{sDP}}$ (red dashed arrow) come close to each other. For both the cuts 6 and 8, the distance betwen the bands is larger. Together with the cuts 1-5, this shows that the bands are the closest to each other at the locations indicated with black arrows in Fig.~1(a)(ii). Altogether, cuts 1-8 suggest that constant-energy cuts both above and below $\epsilon_{\mathrm{sDP}}$ form circular band features with their size shrinking towards the secondary Dirac points. The photointensity is modulated along these circular contours [see Fig.~1(a)(iii)], similarly as it is for the original Dirac points.

\subsection*{Comparison of spectra taken with photons of different energy}

\begin{figure}[!b]
\centering
\includegraphics[scale=1.0]{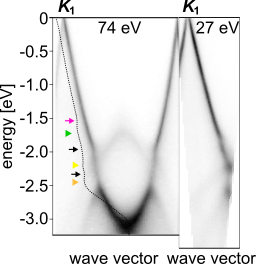}
\caption{Comparison of $k$-space cuts along the direction $\vect{K}_{1}$ to $\vect{K}_{1}^{'}$ taken with photons with energy 74 eV and 27 eV.}
\label{fig:S5}
\end{figure}

In Fig.~\ref{fig:S5}, we compare photoemission spectra along the direction $\vect{K}_{1}$ to $\vect{K}_{1}^{'}$ for tBLG-C taken with photons with energies 74 eV and 27 eV [the panel for 74 eV photons is the panel shown in the main text in Fig.~3(b)]. While for the lower energy photons we imaged a shorter distance in the reciprocal space, it is clear that the minigap features close to $\vect{K}_{1}$ are present in both spectra. Also, no new features appear at 27 eV and similar behaviour of photointensity for both energies contributes to us excluding final state effects, rather than minigaps, as a feasible explanation for our observations.   

\section*{Theoretical model of minibands and photoemission spectra}

\subsection*{Electronic spectrum of twisted bilayer graphene}

We describe twisted bilayer graphene using the Hamiltonian \cite{bistritzer_pnas_2011},
\begin{align} \label{ttlg}
\op{H} & = \begin{pmatrix}
\op{H}_{0}\left(\tfrac{\theta}{2}\right) & \op{T}(\theta) \\
\op{T}^\dagger(\theta) & \op{H}_{0}\left(-\tfrac{\theta}{2}\right)
\end{pmatrix}, \\
\op{H}_{0}(\theta) & =
\begin{pmatrix}
0 & -\gamma_0 f(\op{R}_{\theta}\vect{k})\\ 
-\gamma_0 f^*(\op{R}_{\theta}\vect{k}) & 0
\end{pmatrix}, \nonumber  \\
f(\vect{k}) & = \exp \left( \dfrac{i k_y a}{\sqrt{3}}\right) + 2 \exp \left(-\dfrac{i k_y a}{2 \sqrt{3}}\right) \cos\left(\dfrac{k_x a}{2}\right),\nonumber
\end{align}
written in the basis of the sublattice states constructed of carbon $p_{z}$ orbitals $\phi(\vect{r},z)\equiv\phi(x,y,z)$, 
\begin{align}\label{eqn:Bloch}
\Ket{\vect{k},X}_l = \frac{1}{\sqrt{N}}\sum_{\vect{R}_{l}} e^{i\vect{k}\cdot(\vect{R}_{l}+\vect{\tau}_{X,l})} \phi(\vect{r}-\vect{R}_{l}-\vect{\tau}_{X,l},z-z_{l}),
\end{align}
where $\op{R}_{\theta}$ is an operator of clockwise rotation, $a$ is the graphene lattice constant, $\vect{k}=(k_{x},k_{y})$ is electron wave vector, $X=A,B$ is the sublattice, $\vect{R}_{l}$ are the lattice vectors of layer $l$, $\vect{\tau}_{X,l}$ points to the site $X$ in layer $l$ within the unit cell selected by $\vect{R}_{l}$ and $z_{l}$ defines the position of layer $l$ along the $z$-axis. The diagonal blocks, $\op{H}_{0}$, describe electrons in each of the graphene layers based on the standard tight-binding approach \cite{castro_neto_rmp_2009}. We write the interlayer coupling following earlier work \cite{bistritzer_pnas_2011, koshino_njp_2015},
\begin{align}
_{1}\!\braket{\vect{k}',m,|\op{T}|\vect{k},j}_{2} = \sum_{\vect{G},\vect{G}'}\tilde{t}(\vect{k}+\vect{G},c_{0})e^{-i\vect{G}\cdot\vect{\tau}_{j,2}}e^{i\vect{G}'\cdot \op{R}_\theta \vect{\tau}_{m,1}}\,\delta_{\vect{k}+\vect{G},\vect{k}'+\vect{G}'}.
\end{align}
The matrix element above can be interpreted in the following way: (1) the Kronecker delta term, $\delta_{\vect{k}+\vect{G},\vect{k}'+\vect{G}'}$, expresses conservation of crystal momentum and determines the momenta on the top and bottom layers which are coupled (momenta $\vect{k}$ and $\vect{k}'$ offset by a moir\'{e} reciprocal lattice vector $\vect{g}=\vect{G}'-\vect{G}$); (2) the phase  $ e^{i (\vect{G}'\cdot \op{R}_\theta \vect{\tau}_{m,1}-\vect{G} \cdot \vect{\tau}_{l,2})}$ is a phase factor associated with the coupling of orbitals $m$ and $l$ as a result of translations by reciprocal lattice vectors in each layer. Qualitatively, these phases describe a continuous transition between regions of $AA$, $AB$ and $BA$-like stacking present in the moir\'{e} pattern. Finally, the strength of the coupling, $\tilde{t}(\vect{k}+\vect{G}, c_0)$, is prescribed by the (total) momentum of the electron tunnelling between the layers. Here, we use parametrization of $\tilde{t}(\vect{k}+\vect{G}, c_0)$ as described in Ref.~[7] but with $V^{0}_{pp\pi}=2.9$ eV to match closer the slope of the Dirac cones. By writing all four matrix elements in the form of a $2\times 2$ matrix, we obtain the interlayer coupling block at the twisted interface, $\op{T}(\theta)$,
\begin{align}\label{eqn:twist}
 \op{T}(\theta) = &\sum_{\vect{G}, \vect{G'}}\tilde{t}(\vect{k}+\vect{G},c_0) \times \begin{pmatrix}
 e^{i \vect{G} \cdot \vect{\tau}} &e^{i (\vect{G} +\op{R}_{\theta}\vect{G'}) \cdot \vect{\tau} } \\
 1 & e^{i \op{R}_{\theta}\vect{G'} \cdot \vect{\tau}}
 \end{pmatrix} \delta_{\vect{k}+\vect{G},\vect{k'}+\vect{G'}},
\end{align}
where $\vect{\tau} = -(0,a/\sqrt{3})$.

In order to obtain the energy dispersion for a given $\vect{k}$, we include into our Hamiltonian states coupled to $\vect{k}$ by the moir\'{e} reciprocal vectors which are less than a distance $\tfrac{28\pi}{3\sqrt{3}r_{AB}}\sin\tfrac{\theta}{2}$ away from it, compute the matrix elements of $\op{H}$ in this truncated basis and diagonalize the resulting matrix numerically. This approach allows to study folding of the graphene dispersions into the moir\'{e} Brillouin zone and can be used for any twist angle. However, some caution is required when studying large twists: for given $\theta$, lack of a matching commensurate structure means that the eigenvalues of $\op{H}$ cannot be immediately interpreted as representing the electronic band structure as they might not be periodic with respect to the moir\'{e} reciprocal vectors. This is not a problem for ARPES simulations because these focus on the momentum-resolved spectral function which is negligible for states folded from distant parts of reciprocal space. 

Note that for $\theta=30^{\circ}$, subtraction of reciprocal vectors of graphene layers from each other generates two sets of moir\'{e} reciprocal vectors and two effective moir\'{e} Brillouin zones of the same size but rotated by 30 degrees. Our theoretical model produces the same miniband and ARPES spectra irrespectively of which configuration is used. However, we do not reproduce the experimental results observed for tBLG-C if we explicitly assume twelve-fold rotational symmetry for $\theta=30^{\circ}$.

\subsection*{Estimate of moir\'{e} scattering gap size}

\begin{figure}[!b]
\centering
\includegraphics[width=1.0\columnwidth]{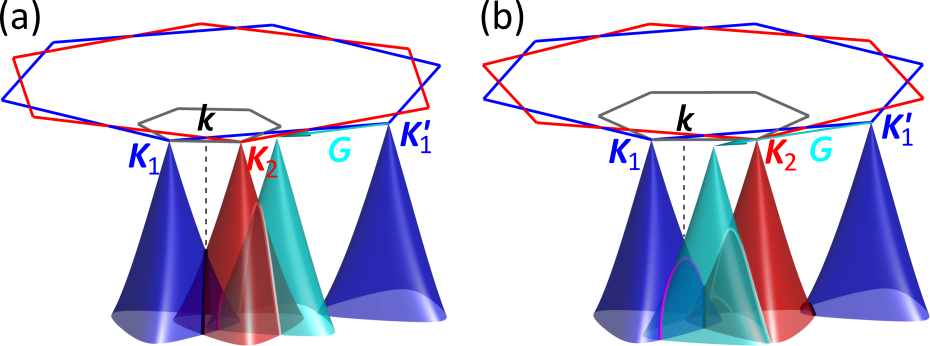}
\caption{Schematic of crossings between unperturbed dispersions of the top (blue) and bottom (red) graphene layers for (a) 'small' twist angle $\theta=18^\circ$, (b) large twist of $\theta=26.5^\circ$. The blue and red hexagons are the BZ of the top and bottom MLG and the corresponding valence band structures in the vicinity of $\vect{K}_{1}$, $\vect{K}^{'}_{1}$ and $\vect{K}_{2}$ are shown with blue and red surfaces. The cyan cone depicts the $\vect{K}^{'}_{1}$ states shifted by a moir\'{e} reciprocal vector indicated with the cyan arrow (the moir\'{e} BZ is shown in gray). Crossings between MLG dispersions are highlighted in black (between two MLG dispersions twisted by $\theta$), magenta (original top MLG dispersion and that translated by a moir\'{e} reciprocal vector) and white (bottom MLG and top MLG translated by a moir\'{e} reciprocal vector).}
\label{fig:S6}
\end{figure}

To estimate the magnitudes of the hybridization and moir\'{e}-induced gaps observed in ARPES spectra, we use a simplified model capturing coupling between the relevant states on two cones of one layer (shown in blue in Fig.~\ref{fig:S6}) and one cone of the second layer (red in Fig.~\ref{fig:S6}),
\begin{align}\label{eqn:3levelH}
\Heff= 
\begin{pmatrix}
\epsilon^{(1)}(\vect{k}) & t & 0 \\
t^{*} & \epsilon^{(2)}(\vect{k}) & \tilde{t}^{*} \\
0 & \tilde{t} & \epsilon^{(1)}(\vect{k}-\vect{G})
\end{pmatrix}.
\end{align}
Above, $\epsilon^{(j)}(\vect{k})$ is the energy of the state in layer $j$ at wave vector $\vect{k}$, $\vect{G}$ is a moir\'{e} reciprocal vector and $t$ and $\tilde{t}$ express couplings between states in different layers (only such coupling is allowed by the interlayer coupling) which we here treat in full generality as a complex number.

We first discuss the standard case of twist angles below our critical angle, $\theta\ll\theta_{c}\approx 21.8^{\circ}$, shown in Fig.~\ref{fig:S6}(a). In this case, the relevant crossing closest to the Dirac points is the one between the blue and red cones. At the wave vector $\vect{k}$ corresponding to the turning point of this crossing, the energies $\epsilon^{(1)}(\vect{k})$ and $\epsilon^{(2)}(\vect{k})$ are equal, $\epsilon^{(1)}(\vect{k})=\epsilon^{(2)}(\vect{k})=\epsilon_{0}$. At the same time, $\epsilon^{(1)}(\vect{k}-\vect{G})=\tilde{\epsilon}\ll\epsilon_{0}$ (cyan surface reaches wave vector $\vect{k}$ well below the energies shown in the figures; note that attempt to couple to the blue cone translated by a longer superlattice reciprocal vector, for example $2\vect{G}$, corresponds to a second-order process and results in a much smaller coupling parameter). We obtain an effective Hamiltonian
\begin{align}\label{eqn:caseI}
\op{\tilde{H}}_{\mathrm{I}} = 
\begin{pmatrix}
\epsilon_{0} & t & 0 \\
t^{*} & \epsilon_{0} & \tilde{t}^{*} \\
0 & \tilde{t} & \tilde{\epsilon}
\end{pmatrix}.
\end{align}
Because $|\tilde{\epsilon}|\gg|\epsilon_{0}|$, we can project our Hamiltonian onto the two relevant states in the vicinity of the crossing which produces an effective Hamiltonian
\begin{align}\label{eqn:caseIeff}
\op{H}_{\mathrm{I}}^{\mathrm{eff}} = 
\begin{pmatrix}
\epsilon_{0} & t \\
t^{*} & \epsilon_{0} 
\end{pmatrix} + 
\begin{pmatrix}0 \\ \tilde{t}^{*}\end{pmatrix}\frac{1}{\epsilon_{0}-\tilde{\epsilon}}\begin{pmatrix}0 & \tilde{t}\end{pmatrix}=
\begin{pmatrix}
\epsilon_{0} & t \\
t^{*} & \epsilon_{0}+\frac{|\tilde{t}|^{2}}{\epsilon_{0}-\tilde{\epsilon}}
\end{pmatrix}.
\end{align}
This, in turn, produces energy eigenvalues
\begin{align}
\epsilon = \epsilon_{0} + \frac{|\tilde{t}|^{2}}{2(\epsilon_{0}-\tilde{\epsilon})}\pm\left[ |t|^{2}+\left(  \frac{|\tilde{t}|^{2}}{2(\epsilon_{0}-\tilde{\epsilon})} \right)^{2} \right]^{\frac{1}{2}},
\end{align}
which for $\tfrac{|\tilde{t}|^{2}}{2(\epsilon_{0}-\tilde{\epsilon})}\ll |t|\approx 0.11$ eV (as is the case here) simplify to the result obtained neglecting moir\'{e} scattering, $\epsilon=\epsilon_{0}\pm|t|$ and the hybridization gap $\Delta_{\mathrm{direct}}=2|t|\approx 0.22$ eV. 

In the case of large twist angles, $\theta>\theta_{c}$, crossing between the $\vect{K}_{1}$ cone and $\vect{K}^{'}_{1}$ cone translated by the superlattice reciprocal vector $\vect{G}$ [blue and cyan in Fig.~\ref{fig:S6}(b)] is closer to the Dirac point than the crossing between $\vect{K}_{1}$ and $\vect{K}_{2}$ [blue and red in Fig.~\ref{fig:S6}(b)]. Using the three-level Hamiltonian, Eq.~\eqref{eqn:3levelH}, for the turning point of this crossing, we have $\epsilon^{(1)}(\vect{k})=\epsilon^{(1)}(\vect{k}-\vect{G})=\epsilon_{0}$, $\epsilon^{(2)}(\vect{k})=\tilde{\epsilon}\ll\epsilon_{0}$,
\begin{align}\label{eqn:caseII}
\op{\tilde{H}}_{\mathrm{II}} = 
\begin{pmatrix}
\epsilon_{0} & t & 0 \\
t^{*} & \tilde{\epsilon} & \tilde{t}^{*} \\
0 & \tilde{t} & \epsilon_{0}
\end{pmatrix}.
\end{align}
Notice that in this case, the states that are degenerate at the energy $\epsilon_{0}$ are not coupled directly as they belong to the same layer. Like before, we can produce an effective Hamiltonian,
\begin{align}\label{eqn:caseIIeff}
\op{H}_{\mathrm{II}}^{\mathrm{eff}} = 
\begin{pmatrix}
\epsilon_{0} & 0 \\
0 & \epsilon_{0} 
\end{pmatrix} + 
\begin{pmatrix}t \\ \tilde{t}\end{pmatrix}\frac{1}{\epsilon_{0}-\tilde{\epsilon}}\begin{pmatrix}t^{*} & \tilde{t}^{*}\end{pmatrix}=
\begin{pmatrix}
\epsilon_{0}+\frac{|t|^{2}}{\epsilon_{0}-\tilde{\epsilon}} & t\tilde{t}^{*} \\
t^{*}\tilde{t} & \epsilon_{0}+\frac{|\tilde{t}|^{2}}{\epsilon_{0}-\tilde{\epsilon}}
\end{pmatrix},
\end{align}
which provides energy eigenvalues
\begin{align}
\epsilon=\epsilon_{0},\,\,\,\,\,\,\,\,\epsilon=\epsilon_{0}+\frac{|t|^{2}+|\tilde{t}|^{2}}{\epsilon_{0}-\tilde{\epsilon}}.
\end{align}

For twisted bilayer graphene, we can assume that $|t|\approx|\tilde{t}|\sim 0.11$ eV \cite{bistritzer_pnas_2011}. This gives the moir\'{e} scattering minigap, $\Delta_{\mathrm{moir\acute{e}}}=\tfrac{|t|^{2}+|\tilde{t}|^{2}}{\epsilon_{0}-\tilde{\epsilon}}\sim 0.02$ eV (where we took $\epsilon_{0}-\tilde{\epsilon}\sim1$ eV). For $\Delta_{\mathrm{direct}}$, our earlier estimate remains the same to the first order. An improved estimate requires investigation of electronic states coupled by the moir\'{e} for a given twist angle to include the most relevant ones in our perturbative description.

Our values for $\Delta_{\mathrm{direct}}$ and $\Delta_{\mathrm{moir\acute{e}}}$ compare reasonably well with the gap sizes observed in our ARPES measurements, Fig.~3 of the main text. In Fig.~3(c), the direct hybridization gap, $\Delta_{\mathrm{direct}}\approx 0.25$ eV, is clearly seen at the lowest energies. The Umklapp minigaps visible at higher energies are an order of magnitude smaller, just like our estimates indicate. The slightly larger magnitude of $\Delta_{\mathrm{direct}}$ and presence of more than one Umklapp minigap is due to the fact that for $\theta$ close to $30^{\circ}$ several scattering processes, rather than just one as in our three-state model, contribute at energies close to $\epsilon_{0}$ \cite{moon_prb_2019}.

\subsection*{ARPES simulations}

Using Fermi's golden rule, we write ARPES intensity as \cite{mucha-kruczynski_prb_2008, mucha-kruczynski_prb_2016}
\begin{align}
I \propto  \sum_{i}\left| M_{f,i} \right|^2 \delta (\omega+\varepsilon_{i,\vect{k}}-W-\varepsilon_{\vect{p}_{e}}),
\end{align}
where $M_{f,i}$ is the matrix element describing transition of the electron from the initial state in the crystal in band $i$ to the final state $f$, $\omega$ is the energy of the incident photon, $\varepsilon_{i,\vect{k}}$ is the energy of an electron in the crystal in band $i$ and with wave vector $\vect{k}$, $\varepsilon_{\vect{p}_{e}}$ is the energy of the photoelectron with momentum $\vect{p}_{e}$ and $W$ is the work function of graphene. Within the dipole approximation,
\begin{align}\label{ARPESLM}
M_{f,i}\propto \bra{\mathrm{final}} \op{A}\cdot\op{p} \ket{\vect{k},i}, 
\end{align}
where $\op{A}$ is the vector potential of the incident photon, $\op{p}$ is the momentum operator, $\ket{\mathrm{final}}$ stands for the final state of the photoelectron and $\ket{\vect{k},i}$ denotes the wave function of the electron in the crystal. The latter is a linear combination of the sublattice Bloch states, Eq.~\eqref{eqn:Bloch}, corresponding to wave vectors connected by a superlattice reciprocal vector $\vect{g}=\vect{G}'-\vect{G}$,
\begin{align}
\ket{\vect{k},i}=\sum_{\vect{g}}\sum_{l,X}c_{X,l}^{\vect{g},i}(\vect{k})\ket{\op{R}_{\theta_{l}}(\vect{k}+\vect{g}),X}_{l},
\end{align}
with the coefficients $c_{X,l}^{\vect{g},i}(\vect{k})$ provided by diagonalization of the Hamiltonian $\op{H}$, Eq.~\eqref{ttlg}. Here, $\vect{g}$ is the moir\'{e} reciprocal superlattice vector. We approximate the final state with a plane wave (justified for incident photon energies above 50 eV \cite{gierz_nanolett_2012}) with momentum $\vect{p}_{e}=(\vect{p}_{e}^{\parallel},p_{e}^{\perp})$, so that
\begin{align*}
M_{f,i}\propto \sum_{\vect{g}}\sum_{l,X}c_{X,l}^{\vect{g},i}(\vect{k})\bra{e^{\tfrac{i}{\hbar}\vect{p}_{e}^{\parallel}\cdot\vect{r}}e^{\tfrac{i}{\hbar}p_{e}^{\perp}z}} \op{A}\cdot\op{p} \ket{\op{R}_{\theta_{l}}(\vect{k}+\vect{g}),X}_{l}.
\end{align*} 
As long as the photon energy is constant and we are only interested in imaging states with similar magnitude of momentum, $|\vect{p}|\approx|\vect{K}|$, the effect of the light-matter interaction, $\op{A}\cdot\op{p}$, can be captured by a phase factor $e^{i\varphi_{X,l}}$ \cite{liu_prl_2011, gierz_prb_2011, hwang_prb_2011},
\begin{align*}\begin{split}
M_{f,i} & \propto \sum_{\vect{g}}\sum_{l,X} e^{i\varphi_{X,l}} c_{X,l}^{\vect{g},i}(\vect{k})\braket{e^{\tfrac{i}{\hbar}\vect{p}_{e}^{\parallel}\cdot\vect{r}}e^{\tfrac{i}{\hbar}p_{e}^{\perp}z}|\op{R}_{\theta_{l}}(\vect{k}+\vect{g}),X}_{l} \\
& = \sum_{\vect{g}}\sum_{l,X, \vect{G}_l} e^{i\varphi_{X,l}}c_{X,l}^{\vect{g},i}(\op{R}_{-\theta_l}(\vect{p}_e^\parallel/\hbar+\vect{G}_l) - \vect{g}) e^{i \vect{G}_l \cdot \vect{\tau}_{X,l}}e^{-\tfrac{i}{\hbar}p_{e}^{\perp}z_l} \tilde{\phi}\left(\op{R}_{\theta_{l}}(\vect{k}+\vect{g})- \vect{G}_l, p_e^\perp/\hbar\right),
\end{split}\end{align*}
where
\begin{align*}
\tilde{\phi}\left(\vect{p}_{e}^{\parallel}/\hbar,{p}_{e}^{\perp}/\hbar\right) = \int d\vect{r}\, dz\, e^{-\tfrac{i}{\hbar} \vect{p}_{e}^{\parallel} \cdot \vect{r}} e^{-\tfrac{i}{\hbar} p_e^\perp z} \phi(\vect{r},z),
\end{align*}
is the Fourier transform of the $p_{z}$ orbital $\phi(\vect{r},z)$. Due to the rotational symmetry of the $p_z$ orbital, $\tilde{\phi}\left(\vect{p}_{e}^{\parallel}/\hbar,{p}_{e}^{\perp}/\hbar\right)=\tilde{\phi}\left(|\vect{p}_{e}^{\parallel}/\hbar|,{p}_{e}^{\perp}/\hbar\right)$. Moreover, for the given photon energy, $\omega$, and work function, $W$, we have $p_e^\perp\gg|\vect{p}_e^\parallel|$ so that $\tilde{\phi}\left(\op{R}_{\theta_{l}}(\vect{k}+\vect{g})- \vect{G}_l, p_e^\perp/\hbar\right)$ can be approximated by a constant and dropped. Finally, in this work we only study points for which $\vect{G}_{l}=0$. As a result, 
\begin{align} \label{eq:Matrix4}
I \propto \sum_{i}\left|  \sum_{\vect{g}}\sum_{l,X}c_{X,l}^{\vect{g},i}(\op{R}_{-\theta_l}\vect{p}_e^\parallel - \vect{g}) e^{i\varphi_{X,l}} e^{-\tfrac{i}{\hbar}p_{e}^{\perp} l c_0}  \right|^2   \delta (\omega+\varepsilon_{i,\vect{k}}-W-\varepsilon_{\vect{p}_{e}}),
\end{align}
where we used the fact that $z_l = l c_0$. We combine both phases $\exp({i\varphi_{X,l}})$ and $\exp({-\tfrac{i}{\hbar}p_{e}^{\perp} l c_0})$ into a single factor,
\begin{align}
    e^{i\varphi_{X,l}} e^{-\tfrac{i}{\hbar}p_{e}^{\perp} l c_0} =e^{i\alpha_{X,l}}, 
\end{align}
which we fit to experiment (the experimental data suggests that the phase difference between two neighbouring graphene layers is approximately $e^{i\pi}$).

Finally, we model the Dirac delta in Eq.~\eqref{eq:Matrix4} with a Lorentzian
\begin{align}
\delta( \omega+\varepsilon_{i,\vect{k}}-W-\varepsilon_{\vect{p}_{e}})\rightarrow \dfrac{1}{\pi} \dfrac{\gamma}{( \omega+\varepsilon_{i,\vect{k}}-W-\varepsilon_{\vect{p}_{e}} )^2 + \gamma^2},   
\end{align}
with half-width-half-maximum $\gamma$.

\begin{figure}[!t]
\centering
\includegraphics[scale=1.0]{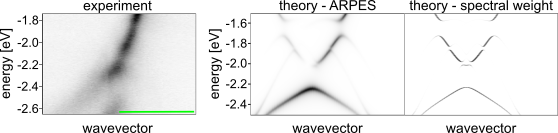}
\caption{Comparison of the experimental and theoretical photoemission spectra from Fig.~3(c) of the main text with $S(\vect{k},\epsilon_{\vect{k},i})$ for the same range of energies and wave vectors as the theoretical ARPES simulation.}
\label{fig:S7}
\end{figure}

Using our knowledge of the electronic wave functions, we can also compute the single particle spectral weight of the wave function of band $i$ at wave vector $\vect{k}$,
\begin{align}
S_{i}(\vect{k},\epsilon_{\vect{k},i})=\sum_{l}|{}{_{\,l}}{\braket{\vect{k}|\vect{k},i}}|^{2},
\end{align}
where $\ket{\vect{k}}_{l}$ is the unperturbed eigenstate of layer $l$ at wave vector $\vect{k}$. In the absence of interlayer coupling, the wave function $\ket{\vect{k},i}$ is identical to one of the unperturbed states $\ket{\vect{k}}_{l}$ and $S_{i}(\vect{k},\epsilon_{\vect{k},i})=1$. In the presence of the coupling, $S_{i}(\vect{k},\epsilon_{\vect{k},i})$ provides us information about the proportion of the original unperturbed states at $\vect{k}$ contributing to the moir\'{e} superlattice wave function $\ket{\vect{k},i}$. In Fig.~\ref{fig:S7}, we compare the experimental and theoretical photoemission spectra from Fig.~3(c) of the main text with $S_{i}(\vect{k},\epsilon_{\vect{k},i})$. Notice that all the minigaps we discuss in the main text are already present in $S_{i}(\vect{k},\epsilon_{\vect{k},i})$ and hence are true minigaps and not photointensity modulation features arising due to Berry phase or final state effects (here, we use $S_{i}(\vect{k},\epsilon_{\vect{k},i})$ to study the minibands because plotting all $\epsilon_{\vect{k},i}$ directly for a general twist angle includes all of the folded and moir\'{e}-coupled bands, most of which are weakly coupled and yet complicate interpretation of the spectrum).

\subsection*{Matching of simulations to experimental data}

In Fig.~3(c) of the main text, the energy windows of the simulation and experimental data are not exactly the same. Similarly, in Fig.~1(a) of the main text, the values of the energies as given correspond to the energies of the experimental cuts while the simulated maps have been selected to display similar features and correspond to an energy within $\sim 200$ meV away from the experimental value. There are several reasons for the mismatch between the simulation and experiment: (i) within our theoretical model, the individual graphene layers are described using a single parameter, the nearest neighbour hopping, $\gamma_{0}$, which makes matching experimental features across the energy range of over 3 eV away from the Dirac points difficult \cite{jung_prb_2013}; (ii) the experimental energy resolution is about 75-80 meV [except the closeup in Fig.~2(c) for which the resolution is $\sim 45$ meV]; (iii) the accuracy with which we can determine the positions of the Dirac points of both layers (which specify experimental zero of energy) is $\sim 40$ meV; (iv) finally, nonuniformity of the detector dispersion introduces nonlinear transformation of the position of the feature as measured by the detector to binding energy which leads to an additional shift of up to $\sim 100$ meV at energies $\sim 2$ eV below the Dirac point as compared to energies close to the Dirac point; this shift depends on the details of a particular measurement (for example, photon energy). Given these uncertainties, we are satisfied with the matching of the features between theory and experiment -- attempts to fine-tune the theoretical model to produce a closer quantitative correspondence would add little to the understanding of the observed phenomena.